# Fast fabrication of $WS_2$/$Bi_2Se_3$ heterostructure for high-performance photodetection

*Fan Li1, Jialin Li1, Junsheng Zheng1, Yuanbiao Tong1, Huanfeng Zhu1,2, Pan Wang1,2, and Linjun Li1,2\**

**Author Information**

Corresponding Author

Linjun Li – 1. State Key Laboratory of Modern Optical Instrumentation, College of Optical Science and Engineering, Hangzhou 310000, R. P. China; Email: lilinjun@zju.edu.cn

2. Intelligent Optics & Photonics Research Center, Jiaxing Research Institute Zhejiang University, Jiaxing 314000, China

Authors

Fan Li - State Key Laboratory of Modern Optical Instrumentation, College of Optical Science and Engineering, Hangzhou 310000, R. P. China

Jialin Li - State Key Laboratory of Modern Optical Instrumentation, College of Optical Science and Engineering, Hangzhou 310000, R. P. China

Junsheng Zheng - State Key Laboratory of Modern Optical Instrumentation, College of Optical Science and Engineering, Hangzhou 310000, R. P. China

Yuanbiao Tong - State Key Laboratory of Modern Optical Instrumentation, College of Optical Science and Engineering, Hangzhou 310000, R. P. China

Huanfeng Zhu – 1. State Key Laboratory of Modern Optical Instrumentation, College of Optical Science and Engineering, Hangzhou 310000, R. P. China

2. Intelligent Optics & Photonics Research Center, Jiaxing Research Institute Zhejiang University, Jiaxing 314000, China

Pan Wang – 1. State Key Laboratory of Modern Optical Instrumentation, College of Optical Science and Engineering, Hangzhou 310000, R. P. China

2. Intelligent Optics & Photonics Research Center, Jiaxing Research Institute Zhejiang University, Jiaxing 314000, China

photodetectors

**Abstract**

Two-dimensional (2D) material heterostructures have attracted considerable attention owing to their interesting and novel physical properties which expand the possibilities for future optoelectronic, photovoltaic and nanoelectronic applications. A portable, fast and deterministic transfer technique is highly needed for the fabrication of heterostructures. Herein, we report a fast half-wet polydimethylsiloxane (PDMS) transfer process utilizing the change of adhesion energy with the help of micron sized water droplets. Using this method, a vertical stacking of $WS_2/Bi_2Se_3$ heterostructure with a straddling band configuration is successfully assembled on fluorophlogopite substrate. Thanks to the complementary bandgaps and high efficiency of interfacial charge-transfer, the photodetector based on the heterostructure exhibits a superior responsivity of 109.9 A $W^{-1}$ for a visible incident light at 473 nm and 26.7 A $W^{-1}$ for a 1064 nm near-infrared illumination. Such high photoresponsivity of the heterostructure demonstrates that our transfer method not only owns time efficiency but also ensures high quality of the heterointerface. Our study may open new pathways to the fast and massive fabrication of various vertical 2D heterostructures for applications in twistronics/valleytronics and other band engineering devices.

**Introduction**

Van der Waals (vdW) heterostructures based on vertically stacked two-dimensional (2D) materials have been widely studied owing to their synergistic role that improving the functional characteristics.[1, 2] New functionalities can arise from the interface between dissimilar 2D materials, or from the differences of their electronic structures, lattice constants and Young's modulus, creating quantum wells, twisted superlattices, strong light-matter interaction and superior optoelectronic properties.[3-6]

There have been extensive reports on methods of producing 2D heterostructures, which can be divided into two directions: in situ direct growth and multistep mechanical transfer.

In situ direct growth means direct growth of different layers sequentially. The general techniques, such as molecular beam epitaxy (MBE), physical vapor deposition (PVD), pulsed laser deposition (PLD), and chemical vapor deposition (CVD) for monolayer growth, can be slightly modified to realize the direct fabrication of heterostructures.[7-10] MBE potentially allows precise control over elemental deposition rates and easy switching from one material to another, enabling the fabrication of the heterostructures on wafer-scales. Several kinds of heterostructures have been reported, like $MoTe_2/MoS_2$, $MoSe_2/HfSe_2$ and $Bi_2Se_3$/graphene, via one-step growth or in situ growth on the bottom material.[11-14] PVD offers many advantages such as selective and large-area growth, high growth rate and low cost. $WS_2/MoS_2$ heterojunction can be assembled via reactive sputtering deposition (RSD), which is one type of PVD.[9] Moreover, PLD is also a type of PVD methods, which is precursor free and provides controllable growth rate and sample thickness via changing the energy fluence of the pulsed laser. A few reports demonstrate this large-area growth technology, $WS_2/Bi_2Te_3$ heterostructure and $WS_2/WSe_2/WS_2$ tri-layer heterostructure have been fabricated and demonstrate great performance in photodetection and quantum well, respectively.[8, 15] CVD is an efficient and scalable method in the preparation of monolayer and bilayer 2D materials, which is also an alternative choice for the fabrication of heterostructures.[16] Gong et al. achieved direct growth of $MoS_2/WS_2$ heterostructures on $SiO_2/Si$ substrate.[7] Several other van der Waals heterostructures, such as $MoS_2/WSe_2$, $MoS_2/CdS$, $WSe_2/SnS_2$ have been successfully synthesized by the CVD method.[17-19] Although in situ growth technology has plenty of advantages, like large scale, considerable productivity and low cost, it is still confronted with certain challenges. Firstly, the constructure of the heterostructures, like the twist angle between two layers, is fixed by the self-assembly process.[10] Furthermore, the random position growth leads to low reproducibility. Secondly, it is hard to satisfy the growth conditions of the two target materials simultaneously, such as growth temperature and gas atmosphere, which limits the categories of heterostructures that in situ direct growth can achieve.[20]

As a result, traditional mechanical transfer methods still have irreplaceable advantages. The pick-drop process has almost no restrictions on the categories of the target samples. In some degree, every two types of thin film materials could be combined and form a heterostructure. Moreover, this technique provides controllable precise stacking alignment, which makes near-zero twisted heterostructures possible.[4] One of the most commonly reported methods is the polymethylmethacrylate (PMMA) film based wet transfer, which uses a PMMA film spin-coated on substrates as an adhesion layer and support.[21] Utilizing the contrast between hydrophilic and hydrophobic of PMMA and growth substrate or using corrosive solvent (like NaOH or HF aqueous solution) to separate substrates and PMMA adhered with target material. The whole PMMA film can be placed onto another target substrate and after dissolving PMMA, the target material remains on the substrate and the transfer finishes.[22] However, the PMMA is hard to be dissolved completely and the residue remaining on the sample degrades its device performance and limits its potential applications. Moreover, the wrinkling of the PMMA film cannot be avoided, which makes precise placement challenging and brings unwanted deformation of the sample. Another general method via the viscidity of polydimethylsiloxane (PDMS) can easily pick up mechanically exfoliated flakes and release the flakes by stronger van der Waals force between flakes and the bottom substrate or sample.[23] Nevertheless, it's hard for PDMS to pick up samples grown by CVD because of the robust adhesion between samples and the rigid grown substrate. Other effective transfer methods have been reported such as using a polypropylene carbonate (PPC) coated PDMS block as the transfer support or polymer support/Graphene/target substrate roll-to roll transfer.[24, 25] These methods either involve residues or have difficulty on picking up CVD grown flakes. Due to current status, there is still urgent need to develop a clean, fast and easy aligning transfer method to fabrication of vdW heterostructures exploiting the massively CVD grown materials.

Herein, we report a fast half-wet PDMS transfer method that utilizes the change of adhesion energy with the help of hot water vapor and compare the complexity of different grown substrates. In previous reports, X. Ma et al. explored the assisted effect of capillary force induced by water vapor and successfully transferred graphene and $MoS_2$ domains grown on $SiO_2$/Si wafer. But they didn't explore the applicability of this method in other substrate, such as sapphire and mica. Also, it lacks fine control of material placement in an in situ manner and more precise transfer manipulation is needed for large scale monolayer.[26] Lee et al. achieved high quality graphene transfer from an $Al_2O_3$ substrate via deionized (DI) water, while Y. Hou et al. utilized similar method to fabricate twisted bilayer graphene.[27, 28] However, they both focused on mechanical exfoliated graphene and a pretreated step was needed, while samples grown by CVD were still neglected. S. Dong et al. expanded the transfer method to transition metal dichalcogenides (TMD) grown by CVD via assistance of hot DI water, which was induced by pipette or directly dipping PDMS/sample in water.[29] In this process, plenty of water was needed, resulting in long time to dry PDMS and large area of monolayer was unable to be transferred because of the wrinkles caused by water. J. Cai et al. also used trapezoid PDMS and water droplets to transfer $WS_2$ flakes, while the size of the transferred flakes was limited to a few microns.[30] Compared to previously reported methods, our method not only combines their advantages, but also explores the applicability to different substrates (such as sapphire and mica) and realizes the transfer of large-area single monolayer flakes. With the assistance of micro sized water droplets instead of spraying or immersion in water, PDMS can be switched from wet to dry in seconds, which promotes the efficiency of transfer and maintaining the original flatness of large monolayer. The whole transfer process can be accompanied in a few minutes.

Furthermore, we fabricate a $WS_2$/$Bi_2Se_3$ heterostructure for photodetection via our new method and CVD/PVD grown flakes. The heterostructure with a type-I band alignment shows an excellent photodetection performance from visible to near-infrared illumination. The response parameters of the resultant device are superior to that of the previous reports. For example, the responsivity of our device is 74.5 $AW^{-1}$ at visible 532 nm light illumination, which is $10^3$ higher than $MoS_2$/$WS_2$ heterostructure, while the responsivity at near-infrared 1064 nm is nearly $2.7\times10^4$ higher than $WS_2$/$Bi_2O_2Se$ device.[10, 31] Besides, the on/off ratio of $1.06\times10^6$ and $7.2\times10^4$ at 473 nm and 1064 nm respectively, is $1.7\times10^3$ higher than that of reported $WSe_2$/$Bi_2O_2Se$ structure.[32] The response speed at visible illumination of our device is 18 times faster than previous $WS_2$/$Bi_2Se_3$ heterostructure grown by PVD.[33] such outstanding performance manifests that our methodology has potential applications in fast and massive fabrications for high quality functional heterostructure devices and twistronics/valleytronics.

**Results and discussion**
**Fast half-wet transfer process**
The heterostructure is assembled by a fast half-wet transfer process as mentioned above. The basic procedure of the transfer process is illustrated in **Figure 1**. The setup is developed consisting of a glass slide mounted on a micromanipulator, a bottom stage with translation capability and an optical microscope for visualization. First, a small piece of PDMS is cut and stuck onto a glass slide for adhesion and support. A flask filled with hot water (80 °C) is placed below PDMS and kept for about 10 seconds. The hot vapor condenses on PDMS and forms micro droplets. Then fresh monolayer $WS_2$ grown on a sapphire substrate is moved to the bottom of PDMS by the translation stage and a suitable $WS_2$ sample is selected by the microscope. After achieving contact between PDMS and the sapphire by lifting off the glass slide, the droplets penetrate into the interface and decrease the adhesion energy of $WS_2$ on the sapphire, which make it easier for PDMS to pick $WS_2$ up. The PDMS moves up with picked $WS_2$, the sapphire is replaced by the target mica with PVD grown $Bi_2Se_3$ flakes. By employing the microscope imaging and 3-axis micromanipulator stage, monolayer $WS_2$ is moved to the aligned location and dropped onto the $Bi_2Se_3$ slowly and softly. Finally, by careful manipulation through the z-motion of the glass slide-carrying stage, PDMS is peeled off from the mica and monolayer $WS_2$ remains on $Bi_2Se_3$ because of van der Waals attracting force. The optical images of the corresponding transfer steps are shown in Figure S1 and S2. In order to tightly combine these two materials, the heterostructure is put into the glove box and annealed at 150 °C for 30 minutes. The final fabrication of $WS_2/Bi_2Se_3$ heterostructure device is completed via standard electron-beam lithography, with the 5/75 nm Cr/Au electrodes deposited by thermal evaporation.

During the transfer process, the micron sized droplets evaporate quickly, making it easier to precisely drop $WS_2$ and avoiding wrinkling of the monolayer flake. As a result, no additional steps to dry PDMS are needed, which not only saves time, but also simplifies manipulation and avoids possible folding of monolayer $WS_2$. In the present setup, only the chosen area of $WS_2$ is taken up by the transfer process, while the rest of the substrate area remains intact and is preserved for next transfer process, leading to efficient utilization of $WS_2$. Overall, we compared the half-wet transfer method to several reported transfer techniques with an emphasis on the optimization of the potential limitations of those methods, as shown in **Table 1**.

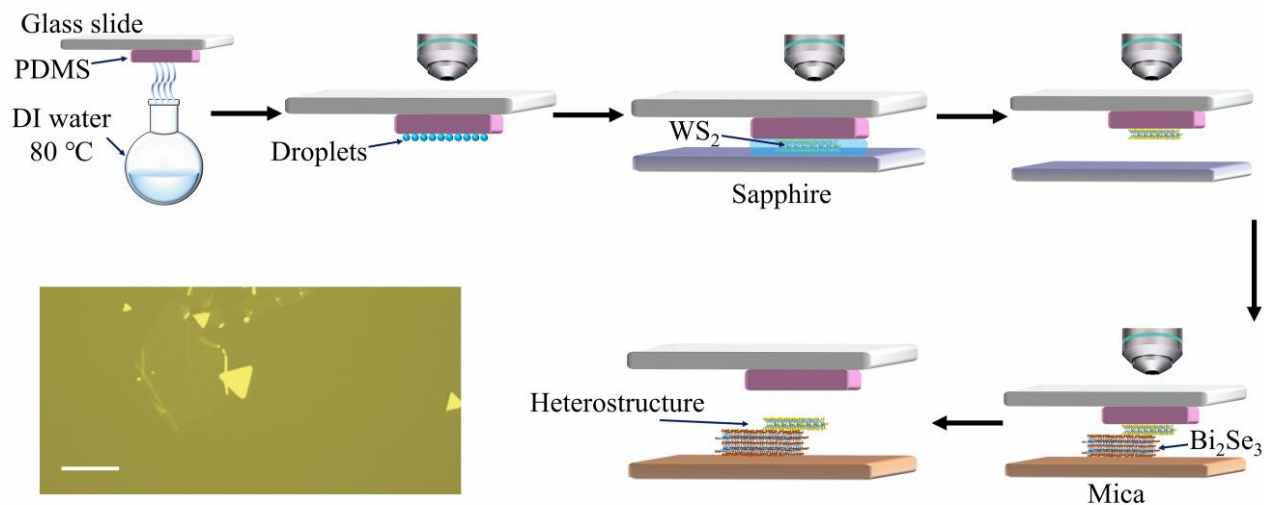

**Figure 1.** Schematic diagram of the transfer and construction processes of WS$_2$/Bi$_2$Se$_3$ heterostructure, inset: optical image of the fabricated heterostructure (scale bar 10 μm).

**Table 1.** Transfer performance comparison of this work with previously reported techniques

| Work | Transfer process | Evaluation | | | |
|---|---|---|---|---|---|
| | | Time consumption | Stacking precision | Sample retaining | Utilization efficiency |
| C. R. Dean et al. [21] Q. Fu et al. [22] | PMMA support film with acetone etching | Overnight soaking to remove PMMA | Hard to locate monolayer on pleated PMMA | Some solvent (like HF) may etch monolayer | Selective transfer of a picked region is not feasible |
| A. C.-Gomez et al. [23] | All-dry viscoelastic stamping via Gelfilm | In a few minutes | Precise alignment via optical microscope | No residual induced | Only for mechanically exfoliated materials |
| F. Pizzocchero et al. [24] | PPC coated PDMS block as transfer support | In a few minutes | Precise alignment via optical microscope | Heating (over 100 ℃) in atmosphere results in oxidation | Only for mechanically exfoliated materials |
| Z. Juang et al. [25] | Roll-to-roll transfer from Ni foil to viscous polymer support film | Rolling speed does not contribute to the transfer results | Not considered | Heating (over 150 ℃) in atmosphere results in oxidation and curling produces folds | Selective transfer of a picked region is not feasible and targeted sample is finite |
| X. Ma et al. [26] | Capillary-force-assisted transfer | In a few minutes | Absence of fine control of sample placement in an in situ manner | No additional pollution | Absence of flexibility in controlling area sizes |
| Lee et al. [27] Y. Hou et al. [28] | Pre-treated substrate and PDMS transfer | In a few minutes | Precise alignment via optical microscope | No additional pollution | Only for mechanically exfoliated materials |

| | | | | | and the size of transferred sample is less than 30 μm |
|---|---|---|---|---|---|
| S. Dong et al.[29] J. Cai et al.[30] | Hot water infiltration assisted PDMS transfer | A certain amount of time needed to immerse and evaporate water | Precise alignment via optical microscope | No additional pollution | The size of transferred materials is a few tens of micrometers |
| This work | Half-wet transfer method | In a few minutes | Precise alignment via optical microscope | No additional pollution | Good versatility for various types of 2D materials grown on various substrates; flexibility in selecting specific samples and the transferred size up to nearly hundred micrometers |

It is worth noting that, compared with other reported transfer results, our half-wet transfer process can pick up single monolayer flake with larger size, up to ~100 μm (shown in Figure S3). A. Kozbial et al. theoretically calculated the surface wettability of $MoS_2$ on $SiO_2$/Si and found the changes in hydrophilicity of $MoS_2$ with hydrocarbon absorption.[34] S. Dong et al. compared the surface energies of TMDCs and $SiO_2$ exposed to water with ambient atmosphere and found water-driven separation of TMDC-$SiO_2$ more energetically favorable.[29] J. Cai et al. proposed the water transport model between two surfaces based on the Young-Laplace equation.[30] The mechanism of water-assisted transfer should have been unambiguous and universal. However, the results based on previous theories showed limited transfer area, which is inconsistent with ours, meaning extra analysis is needed. We propose that the intercalation of water molecule between $WS_2$ and sapphire plays an important role in changing adhesion energy. Though direct evidence can be hardly derived for the intercalating process in current stage, the size of water molecule and crystal structure of $WS_2$ make it comprehensible. According to the previous reported work by Y. Gao et al., the crystal structure of monolayer $WS_2$ was shown via high-resolution transmission electron microscopy (HRTEM) and the distance between W and S atom is about 3 Å, which form periodic hexagons.[35] The size and geometry of water molecule has been widely studied and the length of 2.7 Å is convictive.[36] These results suggest that water molecule is small enough to go through atomic gap of crystal lattice of $WS_2$.[37] After water intercalation and wetting the WS₂/sapphire interface, a large scale monolayer flake can be picked up and the picking speed does not change with the size of transferred flake. Furthermore, we apply half-wet transfer method for different growth substrates, including SiO$_2$/Si, sapphire and mica, and find that the successful transfer rate for the samples on sapphire is the highest, up to almost 100%.

**Characterization of WS$_2$/Bi$_2$Se$_3$ Heterostructure**

**Figure 2**a demonstrates the 3D schematic structure of the WS$_2$/Bi$_2$Se$_3$ heterostructure for photodetection, and an optical microscopic image of the device is provided in Figure 2b. Atomic force microscopy (AFM) confirms the thickness across the heterostructure and each individual flake region as shown in Figure S4. The thickness line profile implies 1 nm thickness of WS$_2$ and 8 nm thickness of Bi$_2$Se$_3$. With a theoretical thickness of monolayer WS$_2$ of 0.7 nm and monolayer Bi$_2$Se$_3$ of 0.955 nm, this result confirms that our WS$_2$ flake is monolayer and Bi$_2$Se$_3$ flake is about 8-layers.[38, 39] In order to confirm the interfacial coupling effect, room temperature photoluminescence (PL) spectra and mapping were collected from the WS$_2$/Bi$_2$Se$_3$ heterostructure (Figure 2c, d). The results indicate that the WS$_2$ exhibits a strong PL emission peak at 2.01 eV (616 nm), which is comparable to a previously reported value of monolayer WS$_2$.[40] The heterostructure region demonstrates an apparent PL quenching effect, as the PL intensity of WS$_2$ decays to 21% of the pristine and remains only 7% compared with the original after annealing at 150 °C. The evolution of PL change as a function of annealing temperature is shown in Figure S5. This suggests the photoexcited carriers are primarily transferred between the WS$_2$ and Bi$_2$Se$_3$ interface, rather than emitting from the individual WS$_2$ band edges.[41] For the Bi$_2$Se$_3$, long-time exposure to air makes its surface oxidation, resulting in the band bending and the change of photodetection performance. Utilizing our fast half-wet transfer method, the time for heterostructure stacking is shortened, just a few minutes. As a result, air exposure time for the whole period of device fabrication can be restricted within 4 hours, meaning that the heterostructure maintains natural properties. Raman spectra of Bi$_2$Se$_3$ after fabricating demonstrates three distinct Raman peak positions at 69, 126, and 173 cm$^{-1}$, corresponding to the $A_{1g}^1$, $E_g^2$ and $A_{1g}^2$ modes, respectively, which are in accordance with the previous reports (Figure 2e).[14] The in-plane and out-of-plane vibration modes for the individual and overlap

region of WS$_2$ at 355 and 420 cm$^{-1}$, respectively, are also shown in Figure 2e. The ratio of the intensities of the A$_{1g}$ and E$_{2g}^1$ peaks is less than 0.5, which also indicates its monolayer thickness.[42] The WS$_2$ and Bi$_2$Se$_3$ characteristic peaks show no shift in the heterostructure, demonstrating that the lattice property of WS$_2$ and Bi$_2$Se$_3$ are not changed after half-wet transfer and thermal annealing.

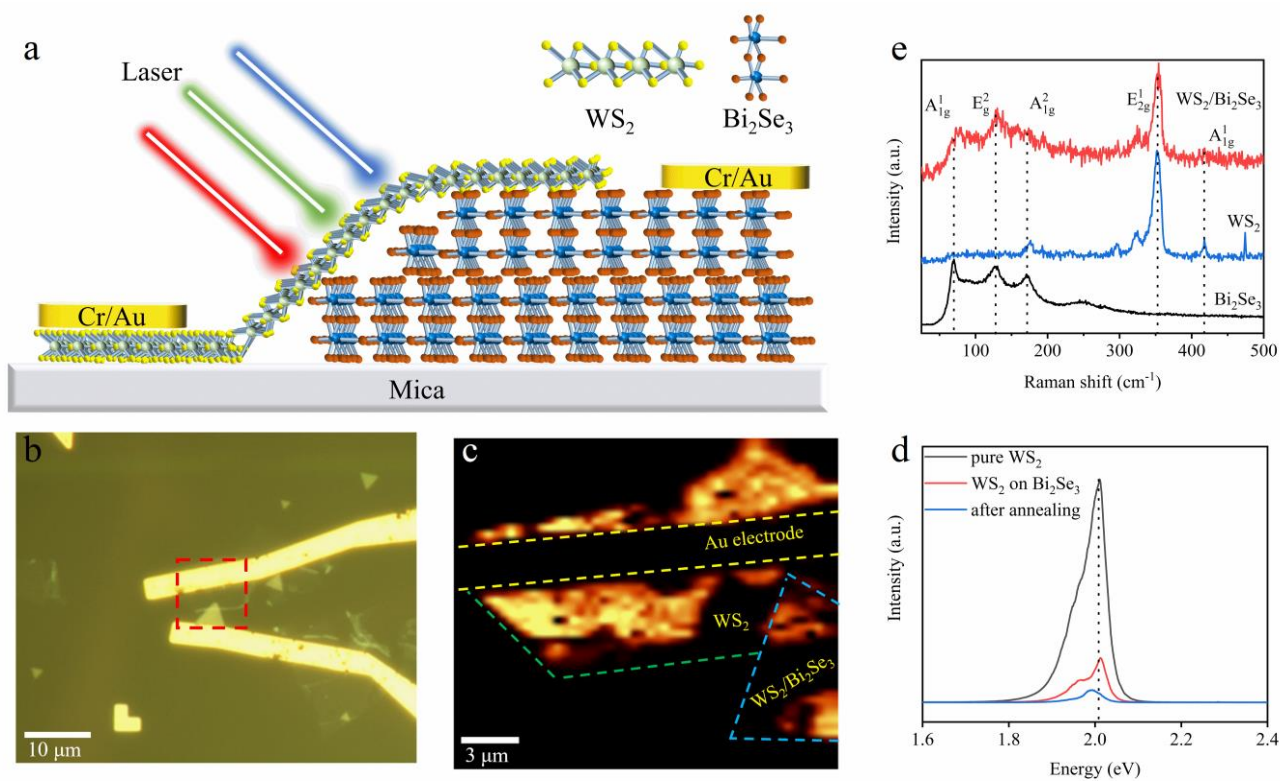

**Figure 2.** Device structure and optical characterizations. (a) Schematic diagram and (b) Optical image of WS$_2$/Bi$_2$Se$_3$ heterostructure device. The monolayer WS$_2$ is stacked on the top of the Bi$_2$Se$_3$ flake. (c) PL mapping of the region surrounded by red dashed lines in Figure 2b. (d) PL spectra measured for WS$_2$ and the heterostructure. (e) Raman spectra of individual Bi$_2$Se$_3$ and WS$_2$/Bi$_2$Se$_3$ heterostructure after device fabrication.

**Electronic Characteristics of WS$_2$/Bi$_2$Se$_3$ Heterostructure**

The electric transport characteristics of the heterostructure device were studied to understand the interaction in the overlapped heterojunction region. All results are acquired by applying a drain voltage ($V_{ds}$) on Bi$_2$Se$_3$ with WS$_2$ being ground. **Figure 3**a presents the current-voltage (*I-V*) curves of the WS$_2$/Bi$_2$Se$_3$ heterostructure in logarithmic (black) and linear (red) scale with the source/drain bias range of -4 to 4 V in the dark. This device shows a rectification behavior with a calculated rectification factor reaching up to 250. To comprehend the carrier transport behavior of the device, the electronic band alignment of the WS$_2$/Bi$_2$Se$_3$ heterostructure is established as shown in Figure 3b. According to the previously reported measurement results and our PL spectrum, the conduction band minimum ($E_c$), valence band maximum ($E_v$) and Fermi level ($E_f$) of the WS$_2$ (Bi$_2$Se$_3$) locate at ~-3.87 eV (-4.91 eV), -5.88 eV (-5.21 eV) and -4.81 eV (-4.61 eV), respectively.[42, 43] From the electronic band before contact, the Fermi level of Bi$_2$Se$_3$ is higher than that of WS$_2$. To prove the band difference, Kelvin probe force microscopy (KPFM) is performed to measure the surface potential of the heterostructure, as shown in Figure S6. Since larger surface potential corresponds to lower work function and thus higher Fermi level, the surface potential of Bi$_2$Se$_3$ is higher than that of WS$_2$ by 0.2 eV, which proves the difference of Fermi level between WS$_2$ and Bi$_2$Se$_3$.[44] Under the reverse bias condition, the external bias voltage breaks the unified Fermi level, elevating the energy band on Bi$_2$Se$_3$ side and descending the WS$_2$ side as shown in Figure 3c. When a small negative $V_{ds}$ is applied, the electrons in the Bi$_2$Se$_3$ are prevented to drift or tunnel to the conduction band of WS$_2$ because of the existence of potential barrier. With the increasing of bias, the energy band of WS$_2$ is further descended, the electrons have the possibility to directly tunnel to WS$_2$. Nevertheless, the amount of tunneling electrons is limited due to the degenerated Bi$_2$Se$_3$ and the light doped of WS$_2$. Tunneling current cannot be significantly higher than the thermionic current, which resulting in a stably low bias current $\approx 6 \times 10^{-12}$ A at -4 V condition In opposite, the energy band of WS$_2$ is elevated under forward bias voltage as shown in Figure 3d, which allows the electrons transfer freely from WS$_2$ to Bi$_2$Se$_3$, resulting in a relatively high current.[45] Because of the unavoidable surface oxidation of Bi$_2$Se$_3$, there exists a barrier between the interface, leading the "open" voltage slightly shifts. Above all, the WS$_2$/Bi$_2$Se$_3$ heterostructure device with a type-I band alignment configuration shows a diode characteristics obviously, indicating that our transfer method ensures high quality of the heterointerface.

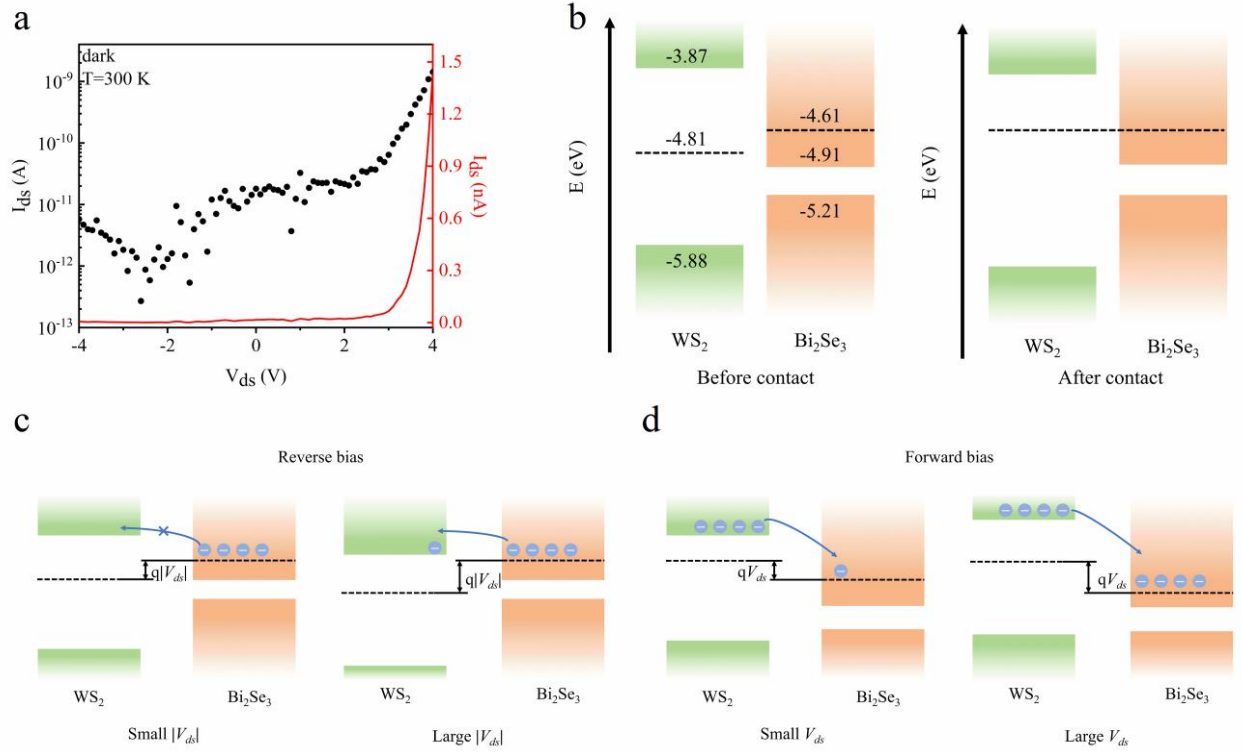

**Figure 3.** Electronic characteristics of WS$_2$/Bi$_2$Se$_3$ heterostructure. (a) *I-V* curves of the heterostructure in dark. (b) Energy band diagram of WS$_2$/Bi$_2$Se$_3$ heterostructure before contact and after contact. (c) and (d) Energy band alignment of WS$_2$/Bi$_2$Se$_3$ heterostructure under different bias voltage $V_{ds}$.

**Photoresponse Properties of WS$_2$/Bi$_2$Se$_3$ Heterostructure**

Considering the complementary bandgaps of monolayer WS$_2$ (2.01 eV) and few-layer Bi$_2$Se$_3$ (0.3 eV), their heterostructure may exhibit excellent photoresponse performances in both visible and near-infrared regions.[31, 46] **Figure 4**a shows the *I-V* curves of the device under dark and different wavelength focused laser illumination with fixed incident power (0.5 mW). Under different condition, the WS$_2$/Bi$_2$Se$_3$ heterostructure shows significant response, demonstrating that the device has a potential in terms of broadband photodetection. Figure 4b, c, d present the output characteristics of the device under 473, 532 and 1064 nm laser light with different power density, respectively. The currents in both forward and reverse biases show increasement with the increase of power density under different laser illumination. In order to evaluate the photodetection performance of the device, the responsivity (*R*) and specific detectivity (*D*$^*$) are calculated. The *R* value can be calculated by the equation of $R = I_{ph}/(P \cdot S_{eff})$, where $I_{ph}$, *P* and $S_{eff}$ are the photocurrent, incident power density and effective areas of the device, respectively. $D^*$ can be extracted by $D^* = R \cdot \sqrt{S_{eff}}/\sqrt{2eI_{dark}}$, where $I_{dark}$ represents the dark current.[47] Figure 4e, f demonstrate the calculated $R$ and $D^*$ under different laser illumination at different excitation power density, respectively. For visible light of 473 nm, the responsivity reaches up to 109.9 A W$^{-1}$ and the detectivity can get to $3.2 \times 10^{12}$ Jones at the power density of 4.527 mWcm$^{-2}$ at $V_{ds}$ = 4 V, while the near-infrared responsivity is up to 26.7 AW$^{-1}$ and detectivity is $7.76 \times 10^{11}$ Jones, which is superior to most reported near-infrared photodetectors.[31, 48]

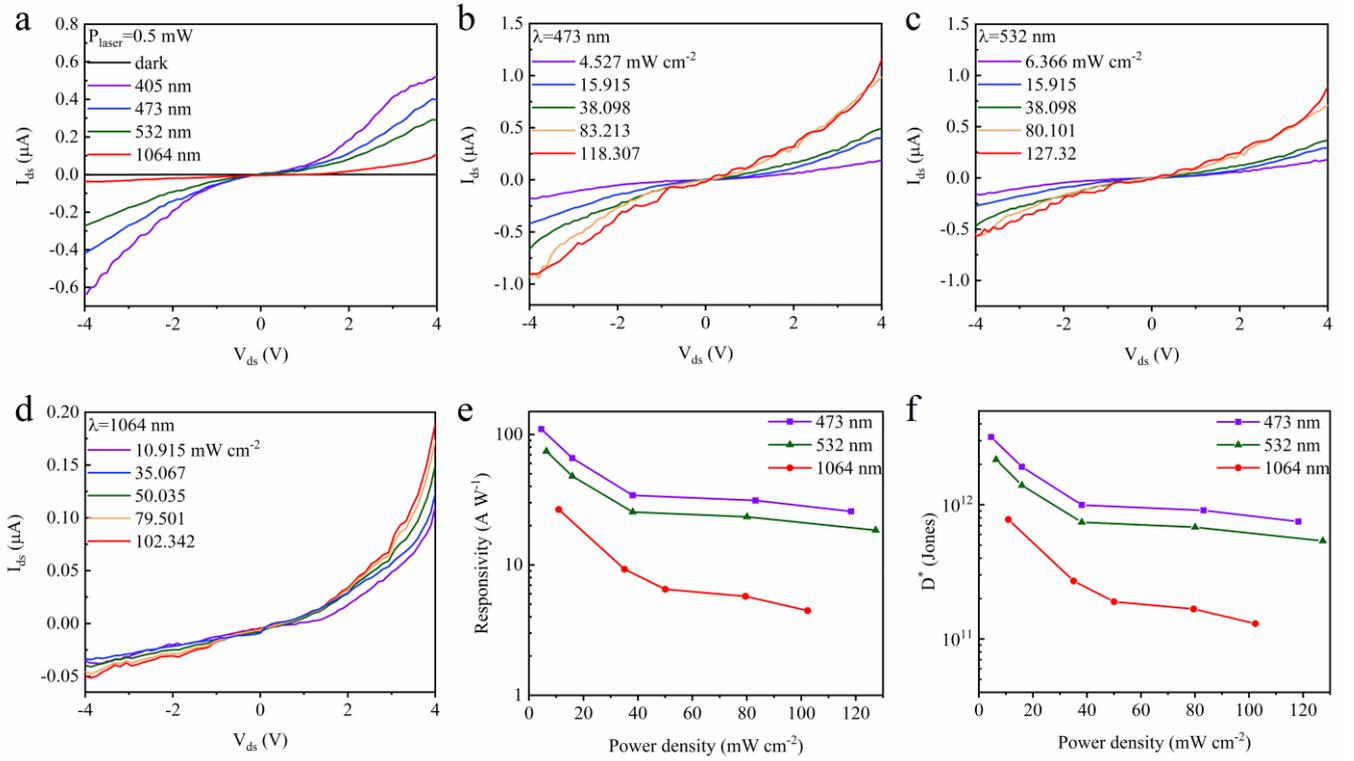

**Figure 4.** Photoelectric characteristics of WS$_2$/Bi$_2$Se$_3$ heterostructure. (a) Broadband response of the device under different wavelength laser illumination at fixed laser power. (b), (c) and (d) Power depended $I_{ds}$-$V_{ds}$ curves under 473, 532 and 1064 nm laser illumination, respectively. (e) and (f) Power depended $R$ and $D^*$ of the device under different wavelength laser illumination at $V_{ds}$=4 V.

To understand the photocurrent generation mechanism, we further studied the response of the device under reverse and forward bias separately. When the device worked at reverse biases voltage, we found that the $I_{ds}$-$V_{ds}$ characteristics could be modeled by the direct tunneling mechanism with linear relationship between $ln(|I_{ds}|/V_{ds}^2)$ and $ln(1/|V_{ds}|)$ as shown in **Figure 5**a. This indicates that the photocarriers generated in the WS$_2$/Bi$_2$Se$_3$ heterostructure are dominated by the tunneling mechanism driven by the external reverse electric field. Under visible laser illumination, both WS$_2$ and Bi$_2$Se$_3$ produce photon-generated electron-hole pairs. The barrier height of the heterojunction region becomes smaller, and the electrons can overcome the barrier or direct tunnel, consequently forming considerable photocurrent (Figure 5b). The carriers are generated only in Bi$_2$Se$_3$ when illuminated by 1064 nm laser as shown in Figure 5c. The photo-generated holes would enter the circuit, while the photo-generated electrons would accumulate at the interface and tunnel into the conduction band of WS$_2$, which exhibits a near-infrared photoresponse. From the linear relationship shown in Figure 5a, a formula like $ln\left(\frac{|I_{ds}|}{V_{ds}^2}\right) = kln\left(\frac{1}{V_{ds}}\right) + b$ can be extracted, where $k$ and $b$ are the slope and y-axial offset of the linear fitting line, which is relating to the probability of tunneling.[49, 50] Then the relationship between photogenerated current and bias voltage can be described as $I_{ds} = e^b \cdot V_{ds}^{2-k}$. From the fitting curves, $k_{visible}$ is smaller than $k_{NIR}$ while $b_{visible}$ is larger than $b_{NIR}$, meaning a higher photogenerated current under visible illumination, which is consistent with the observed phenomena. By fitting the incident power density ($P$) dependent current ($I_{ph}$) through the equation of $I_{ph} \propto P^\alpha$ (Figure 5d), it was found that all the power exponent α values are smaller than 1 under forward bias. This deviation from the ideal value of $\alpha = 1$ suggests a complex process, which relates to the trap states caused by the defects or impurities present in WS$_2$ or Bi$_2$Se$_3$, or the absorbed molecules at WS$_2$/Bi$_2$Se$_3$ interface and slight oxidation on Bi$_2$Se$_3$ surface. These factors can trap the photogenerated charges, so that the absorbed photons cannot be completely converted into photocurrent, resulting in the nonlinear photocurrent response.[51, 52] It can also explain why $R$ and $D^*$ decrease with power density increasing, that is, a drop in the recombination possibility occurs as the photogenerated electrons are captured by the trap states under lower $P$. The transfer processes of visible and near-infrared photo-generated carriers under forward bias are shown in Figure 5e, f. Moreover, the value of external quantum efficiency (EQE) is another parameter of photodetectors, defined by the ratio of the number of electrons collected to the number of incident photons. The calculation and the dependence on power density of EQE are shown in Figure S7.[53] The WS$_2$/Bi$_2$Se$_3$ heterostructure shows an EQE up to 288.6%, corresponding to the photoconductive gain effect caused by interface trapped states.

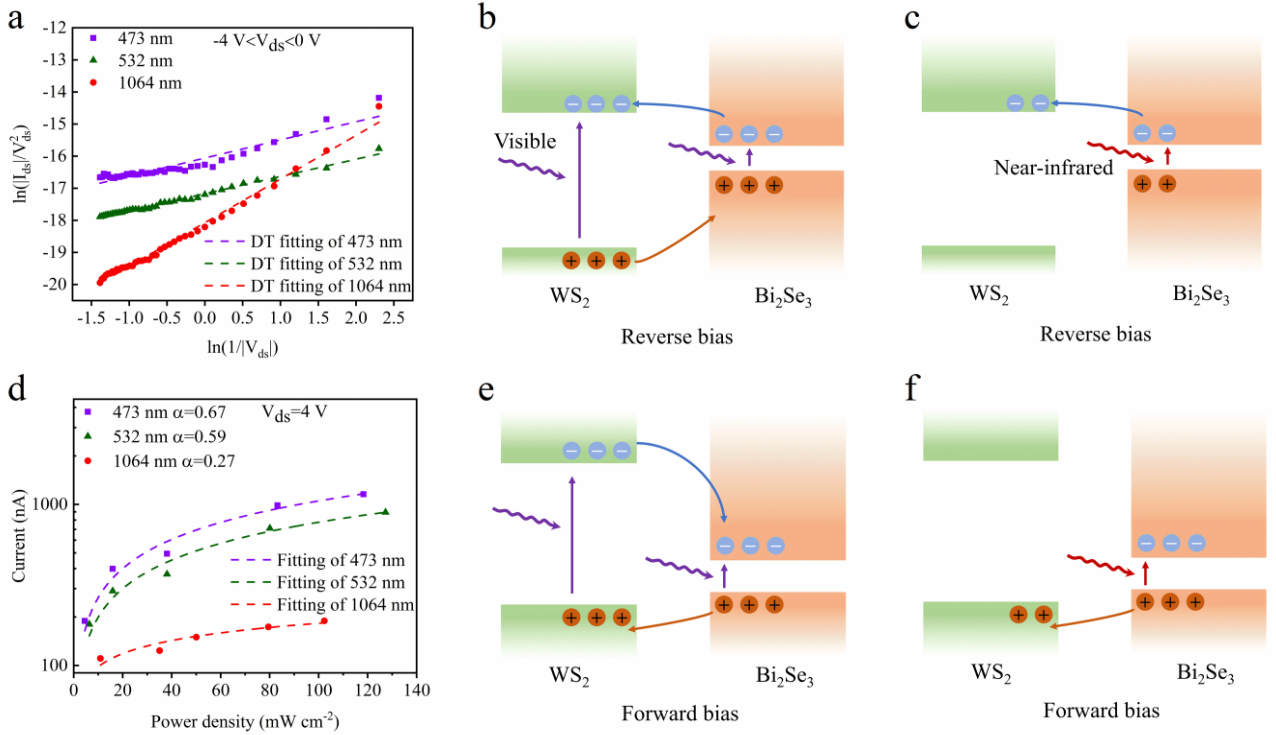

**Figure 5.** Photocurrent generated mechanism of WS$_2$/Bi$_2$Se$_3$ heterostructure. (a) DT fitting of the device at reverse bias voltage. (b) and (c) Energy band diagram of the heterostructure under reverse bias. (d) The photocurrent dependence of laser power density at $V_{ds}$=4 V. (e) and (f) Energy band diagram of the heterostructure under forward bias.

Apart from responsivity, specific detectivity and EQE, response rate and light on/off ratio are also two important parameters for photodetectors. As shown in **Figure 6**a, the WS$_2$/Bi$_2$Se$_3$ heterostructure exhibits well-reproducible photo switching characteristics under different wavelength laser modulation at $V_{ds}$ = 4 V and fixed laser power 0.5 W. Furthermore, the light on/off ratio was obtained by comparing the $I_{ds}$ under illumination and dark. The dependence of light on/off ratio on the bias voltage is shown in Figure 6b. Consequently, we obtained the maximum on/off ratio of ~1.06×10$^6$ for 473 nm @ 118.307 mWcm$^{-2}$, 6.3×10$^5$ for 532 nm @

127.32 mWcm$^{-2}$ and 7.2×10$^4$ for 1064 nm @ 102.342 mWcm$^{-2}$ at $V_{ds}$ = -3 V. In order to evaluate the ability of charge separation and diffusion, Figure 6c and Figure S8 extracts the rise time ($\tau_1$) and fall time ($\tau_2$) of the WS$_2$/Bi$_2$Se$_3$ heterostructure, which are usually defined as the time requirement of the photocurrent increasing from 10% to 90% and decreasing from 90% to 10%, respectively. For 473 nm laser illumination, the rise time and fall time are estimated to be 7 and 3 ms. Moreover, for 1064 nm photoresponse, the rise time of ~13 ms and decay time of ~7 ms are a bit slower than previous reports. The trap state in CVD grown WS$_2$ and slightly unavoidable oxidation on the surface of Bi$_2$Se$_3$ reduce the recombination rate of photo-generated carriers and prolong the life of carriers.[43]

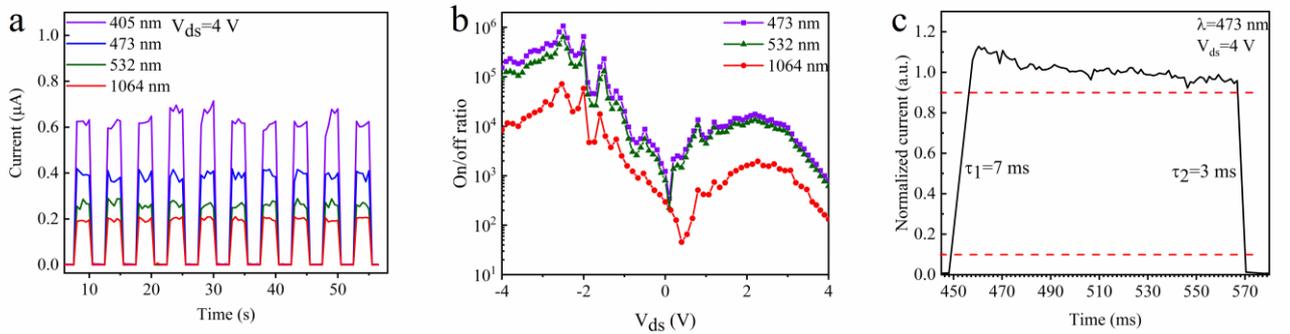

**Figure 6.** Photoresponse characteristics of the WS$_2$/Bi$_2$Se$_3$ heterostructure. (a) Long-cycle stability measurement of the heterostructure device under different wavelength laser illumination. (b) Light on/off ratio as a function of bias voltage under 473 nm@118.307 mW cm$^{-2}$, 532 nm@127.32 mW cm$^{-2}$ and 1064 nm@102.342 mW cm$^{-2}$. (c) Response time curve of the heterostructure under 473 nm laser illumination.

Finally, we make a comparison of our device with similar vdW heterostructures recently reported on the performances of photodetection, as shown in **Table 2**. There are some other photodetection devices we have fabricated performing similarly, as shown in Figure S9 and S10. One can observe that our WS$_2$/Bi$_2$Se$_3$ heterostructure is highly comparable to the state-of-the-art previously reported photodetection materials, suggesting that our transfer method is a promising way for the future fast and massive fabrication of multifunctional heterostructures.

**Table 2**. Photodetection performance comparison of this work with similar devices

| Device structure | Fabrication method | Wavelength [nm] | Bias [V] | Responsivity [A W$^{-1}$] | Detectivity [Jones] | Response rate | On/off ratio |
|---|---|---|---|---|---|---|---|
| MoS$_2$/WS$_2$[10] | CVD growth | 532 | 1 | 0.068 | — | 35/36 ms | — |
| Bi$_2$Se$_3$/graphene[14] | MBE growth | 520 | 1.5 | 1.4 | — | —/114.94 ms | — |
| Bi$_2$O$_2$Se/WS$_2$[31] | PS[b] wet transfer | 532 | 5 | 0.628 | 9.5×10$^8$ | 33/38 ms | 160 |
|  |  | 1350 | 5 | 0.001 | — | — | — |
| Bi$_2$O$_2$Se/WSe$_2$[32] | PVA dry transfer | 635 | 1 | 44 | 3.0×10$^{13}$ | 20/20 ms | 1.2×10$^4$ |
|  |  | 1064 | 1 | 0.01 | 2×10$^8$ | — | — |
| WS$_2$/Bi$_2$Se$_3$[33] | PVD growth | 532 | 1 (V$_g$= -20) | — | — | 126/162 ms | — |
| Bi$_2$Se$_3$/WSe$_2$[43] | PDMS dry transfer | 532 | -5 | 94.26 | 7.9×10$^{12}$ | 1.5/0.11 ms | 7.5×10$^5$ |
|  |  | 1456 | -5 | 3 | 2.2×10$^{10}$ | 4/4 ms | 3.5×10$^4$ |
| Bi$_2$O$_2$Se/BP[48] | PDMS dry transfer | 700 | -1 | 500 | 2.8×10$^{11}$ | —/9 ms | — |
|  |  | 1310 | -1 | 4.3 | 2.4×10$^9$ | — | — |
| WS$_2$[51] | PVA[a] dry transfer | 405 | -2 | 3.45 | — | 31.78/19.81 ms | — |
| WSe$_2$/Bi$_2$O$_2$Se[54] | PDMS dry transfer | 532 | 5 | 443.83 | — | 2.4/2.6 µs | 1.48 |
|  |  |  | -5 | 0.638 | — |  | 618 |
| Bi$_2$O$_2$Se/MoSe$_2$[55] | PMMA wet transfer | 780 | 2 | 0.4131 | 3.7×10$^{11}$ | 0.79/0.49 s | — |
| GaSe/graphene[56] | MBE growth | 532 | -2 | 98 | 3.1×10$^{12}$ | 3/3 ms | 10 |
| WS$_2$/Bi$_2$Se$_3$ (this work) | half-wet transfer | 473 | 4 | 109.9 | 3.2×10$^{12}$ | 7/3 ms | 1.06×10$^6$ |
|  |  | 1064 | 4 | 26.7 | 7.7×10$^{11}$ | 13/7 ms | 7.2×10$^4$ |

[a](poly vinyl alcohol); [b](polystyrene)

## Conclusion

In summary, our study presents a novel 2D material transfer method. By utilizing PDMS and DI water as transfer substrate and lubricant, we successfully realize convenient and fast pick-and-drop process of monolayer WS$_2$. The size of single transferred monolayer piece is up to hundred-micron scale, which is an order of magnitude higher than the original report. To prove the clean and nondestructive transfer process, we fabricate a high quality WS$_2$/Bi$_2$Se$_3$ heterostructure for photodetection. Our heterostructure with type- I band alignment demonstrates wide spectrum response from visible light to near-infrared band (from 405 to 1064 nm). The critical parameters of our heterostructures such as responsivity of 109.9 and 26.7 A $W^{-1}$, light on/off ratio of $1.06\times10^6$ and $7.2\times10^4$ with a fast response time of 7 and 13 ms are obtained at both visible 473 and near-infrared 1064 nm, respectively. Such superior performance demonstrates that our new transfer method may pave a unique avenue to fabricate manifold 2D heterostructures fast and massively.

**Experimental Section**

*Fabrication of $Bi_2Se_3/WS_2$ heterostructure*: First, monolayer $WS_2$ was grown on $Si/SiO_2$ (300 nm) substrates or sapphire (C-plane) by using confined-space chemical vapor deposition at atmospheric pressure.[10] A quartz tube was placed in the center of a horizontal tube furnace (BTF-1200C, Anhui Best Equipment Ltd, China) equipped with an assistant substrate facing upward. The $WO_3$ power was sprinkled on the surface of the assistant substrate while a target growth substrate was placed on the surface of the quartz tube facing down to the assistant substrate. The distance between the assistant substrate and the target growth substrate was about 1 mm. The sulfur power was put into another quartz tube in the center of another furnace upstream. The distance between W and S source was kept as 40 cm. The temperature of W source was first gradually raised up to 600 °C in 60 min, then it was increased to 830 °C with a rate of 10 °C/min and kept at 830 °C for 10 min. When the temperature of W source reached 780 °C, the sulfur power was heated to 200 °C in 5 min and was kept at 200 °C for 10 min for the growth of monolayer $WS_2$. Finally, the tube furnace was naturally cooled down to room temperature. Ar was used as the carrier and protective gas at a flow rate of 50 sccm during the whole growth process.

Then, 2D $Bi_2Se_3$ flakes were grown on flurophlogopite mica substrate by low pressure physical vapor deposition.[45] The $Bi_2Se_3$ power was placed in the center of the furnace equipped with a quartz tube and the flurophlogopite mica substrates were placed in the downstream 8-11 cm away from center. The temperature of $Bi_2Se_3$ was first raised up to 590 °C in 30 min and kept at this temperature for 20 min with Ar carrier gas flow of 100 sccm. The pressure of the furnace was kept at 100 Torr. Finally, the furnace was naturally cooled down to room temperature with a flowing Ar gas.

Next, the monolayer $WS_2$ was transferred onto the $Bi_2Se_3$ flake via a piece of PDMS as the supporting membrane and DI water as lubricant. The transfer details of picking and dropping were discussed in Results and Discussion. The alignment process is shown as follows: We selected one $Bi_2Se_3$ as bottom material and marked its location on the camera screen as we did with $WS_2$. When both $WS_2$ and $Bi_2Se_3$ were settled, by employing the microscope imaging and 3-axis micromanipulator stage, monolayer $WS_2$ was moved to the aligned location and started to drop onto the $Bi_2Se_3$. With the slide going down, we checked the position of $Bi_2Se_3$ from time to time, in order to make sure it still stays at the original position we marked. As distance decreasing, we could see the two materials in the screen till they contacted with each other. Last but not least, by careful manually manipulation through the z-motion of the glass slide-carrying stage, PDMS was peeled off from the mica and monolayer $WS_2$ remained on $Bi_2Se_3$ because of van der Waals force. Finally, electrode patterns were defined by typical electron beam lithography (Nova Nanosem 450) and deposited by thermally evaporating 5/75 nm Cr/Au (K.J.Lesker Nano 36 Thermal Evaporator).

*Characterizations*: Optical microscope (Nikon ECLIPSE LV1SON) and AFM (Park NX-10) were utilized to characterize the morphology of the heterostructure. Raman and PL were investigated by a confocal microscope spectrometer (Alpha300R-2018model, WITech Ltd.) together with 532 laser source. EFM test cantilever (PPP-NCSTAu 3M) was used to measure the surface potential to obtain the difference of Fermi levels. The electronic and photoelectronic measurements of the $Bi_2Se_3/WS_2$ heterostructure were performed by using a Keithley 2450 sourcemeter. Power fixed 405-nm laser and power adjustable 473-, 532- and 1064 nm lasers (Changchun New Industries Optoelectronics Technology Ltd.) were used to record the photoelectric response of the heterostructure. The radius of laser spot in visible waveband was about 1 mm, and the radius of laser spot in 1064 nm was approximately 1.2 mm.

**Supporting Information**

Additional characterization results, including optical image characterization, PL mapping characterization, AFM characterization, KPFM characterization and photodetection performance of additional devices.

**Acknowledgements**

L. J. Li acknowledges the funding from the state key program (91950205), National Key R&D Program of China (2019YFA0308602), and the general program (12174336) of National Science Foundation of China, the Zhejiang Provincial Natural Science Foundation of China (LR20A040002).

**Conflict of Interest**

The authors declare no conflict of interest.

**Data Availability Statement**

The data that support the findings of this study are available from the corresponding author upon reasonable request.

**Fast fabrication of WS$_2$/Bi$_2$Se$_3$ heterostructure for high-performance photodetection**

*Fan Li, Jialin Li, Junsheng Zheng, Yuanbiao Tong, Huanfeng Zhu, Pan Wang, and Linjun Li\**

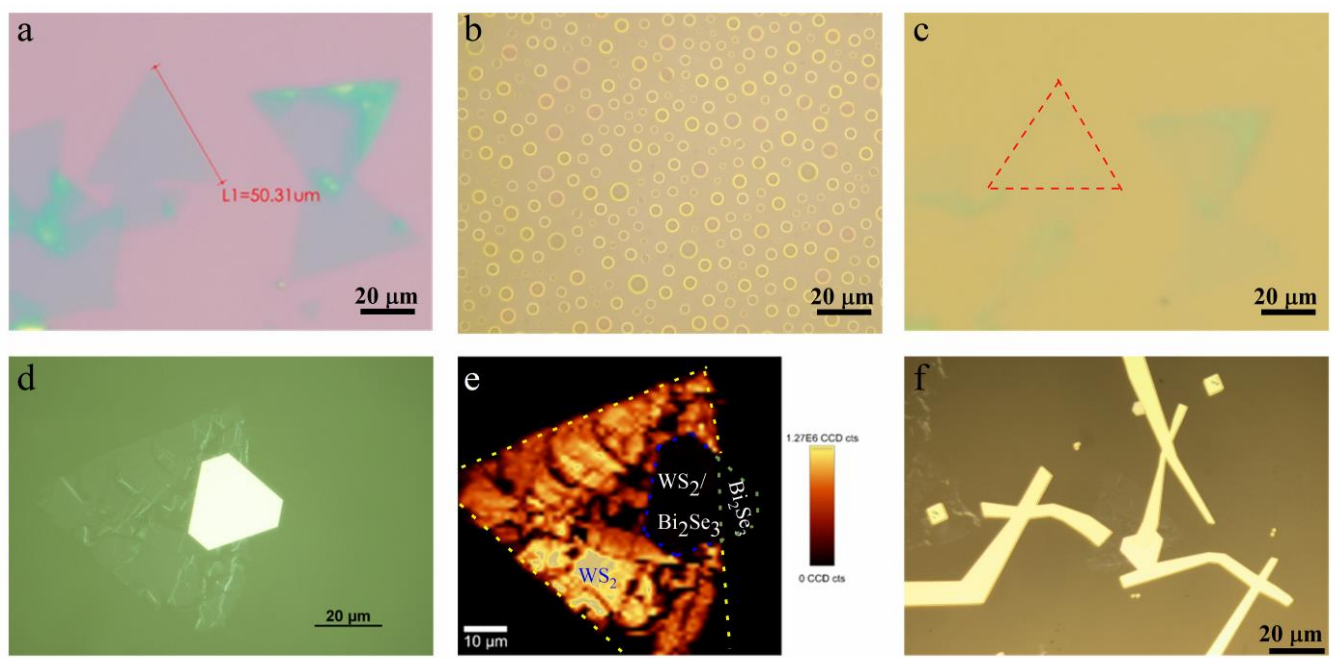

**Figure S1.** Optical image of half-wet transfer process from SiO$_2$/Si to mica. (a) Monolayer WS$_2$ grown on SiO$_2$/Si. (b) The water droplets liquified on PDMS from hot DI water vapour. (c) WS$_2$ lifted on PDMS. (d) The WS$_2$/Bi$_2$Se$_3$ heterostructure. (e) PL mapping of the heterostructure. (f) The photodetection device.

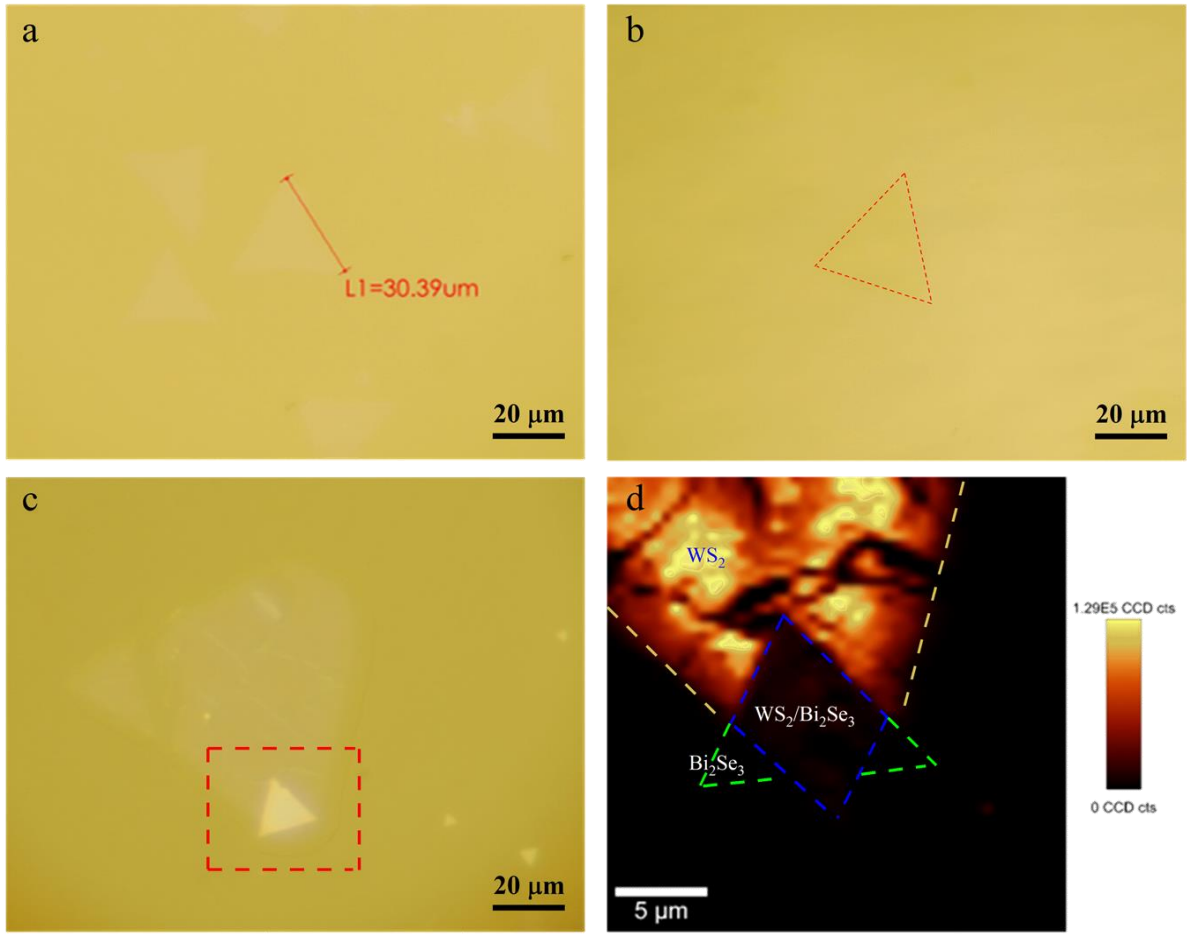

**Figure S2.** Optical image of half-wet transfer process from sapphire to mica. (a) Monolayer $WS_2$ grown on sapphire. (b) $WS_2$ lifted on PDMS. (c) The $WS_2/Bi_2Se_3$ heterostructure. (d) PL mapping of the region surrounded by red dashed lines in (c).

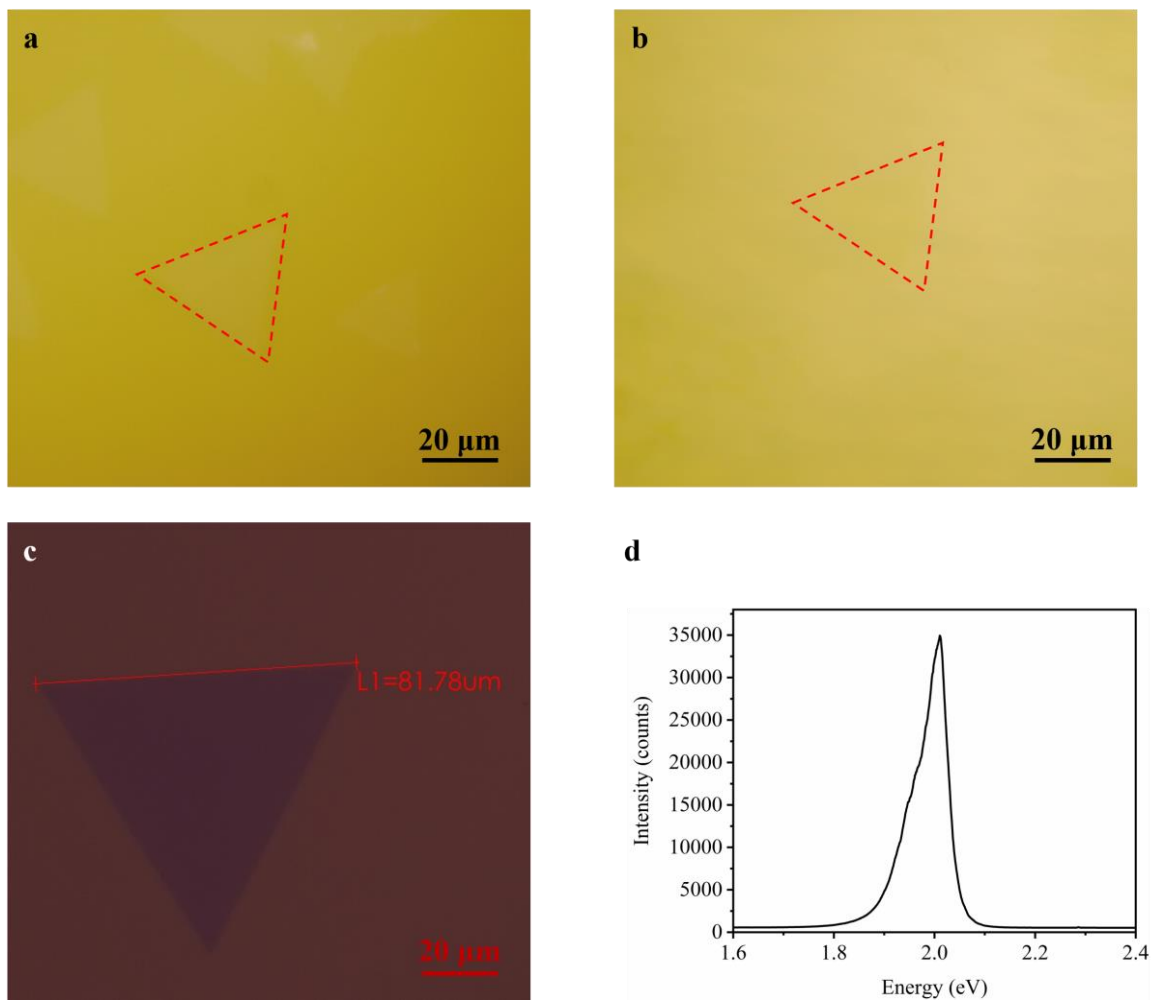

**Figure S3.** Optical image of half-wet transfer process from sapphire to $SiO_2$/Si. (a) Monolayer $WS_2$ grown on sapphire. (b) $WS_2$ lifted on PDMS. (c) $WS_2$ dropped onto $SiO_2$/Si. (d) PL spectra of $WS_2$ with high efficiency.

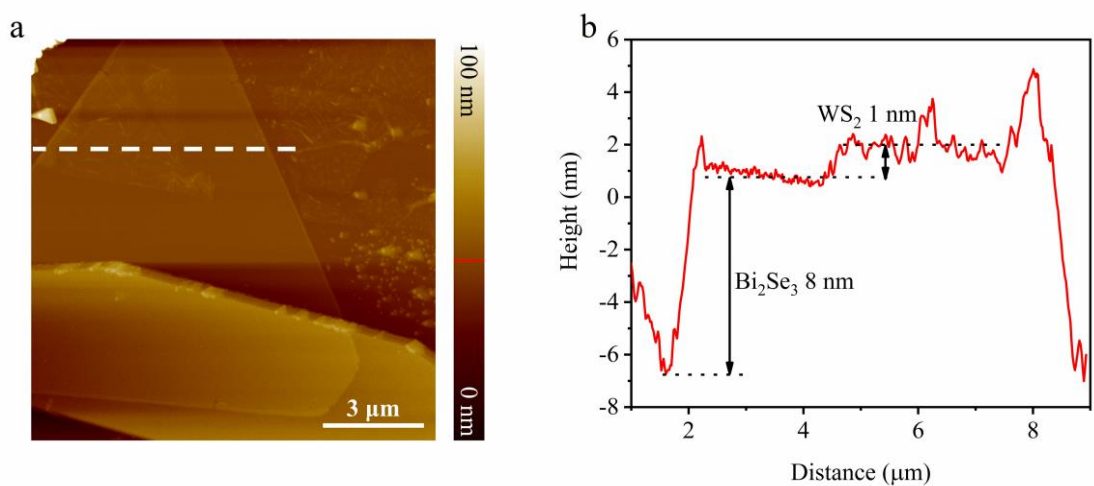

**Figure S4.** AFM of the $WS_2$/$Bi_2Se_3$ heterostructure. (a) AFM mapping of the $WS_2$/$Bi_2Se_3$

heterostructure and individual $Bi_2Se_3$. (b) Height profile of the $WS_2/Bi_2Se_3$ heterostructure in (a) white dotted line.

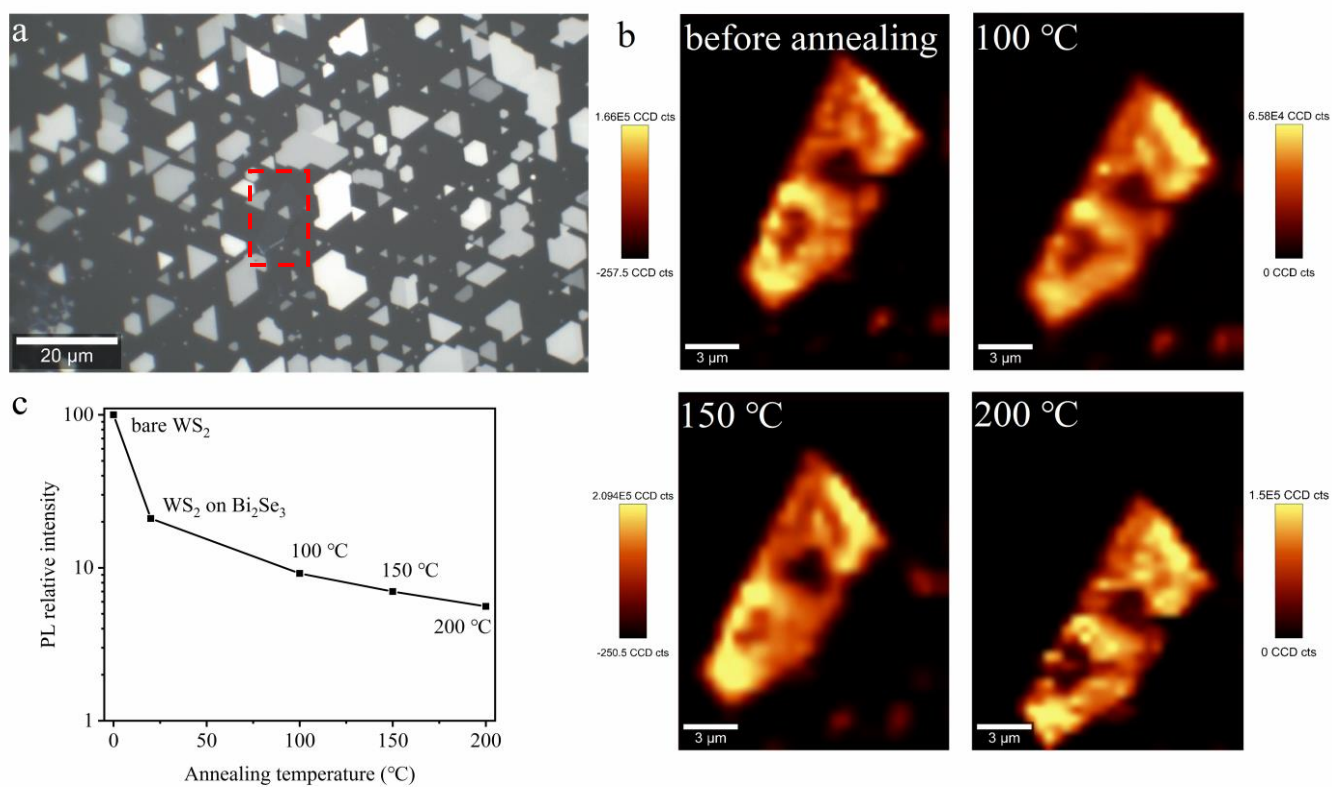

**Figure S5.** The evolution of PL intensity as a function of annealing temperature. (a) Optical image of selected $WS_2/Bi_2Se_3$ heterostructure, marked by red dotted rectangular. (b) PL mapping of selected region after annealing at different temperature. (c) PL relative intensity after annealing at different temperature.

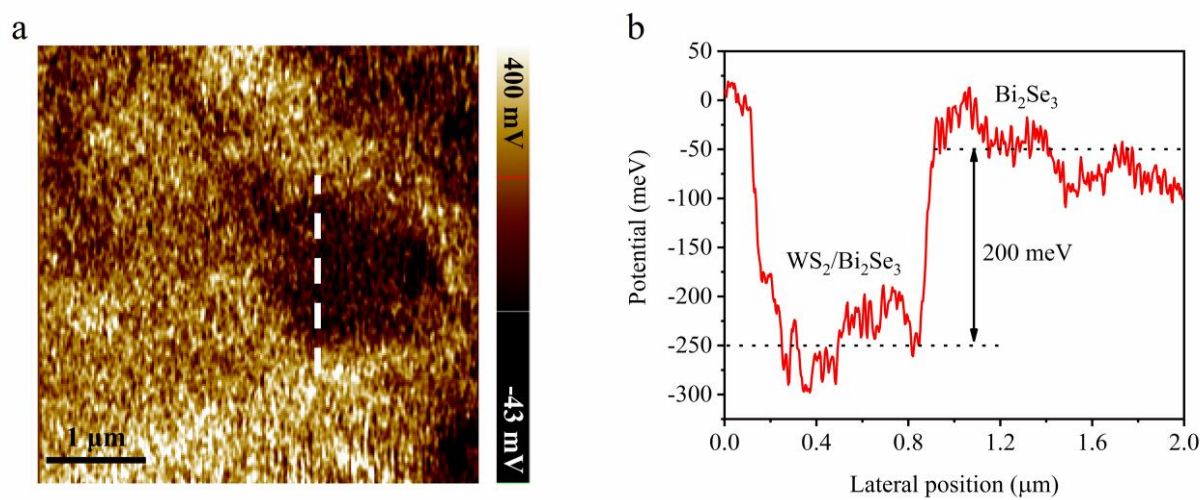

**Figure S6.** KPFM surface potential of the $WS_2/Bi_2Se_3$ heterostructure. (a) The surface potential mapping of the heterostructure and individual $Bi_2Se_3$. (b) The surface potential profile across the sample corresponding to the dashed line in panel (a).

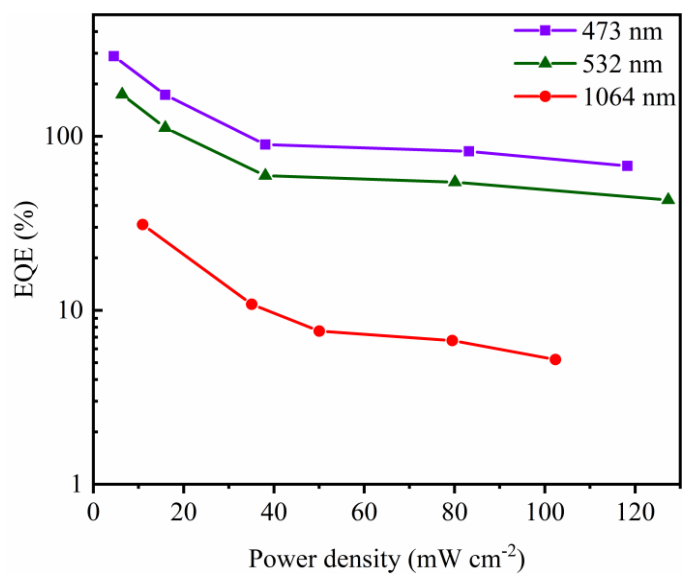

**Figure S7.** EQE of the $WS_2/Bi_2Se_3$ heterostructure device.

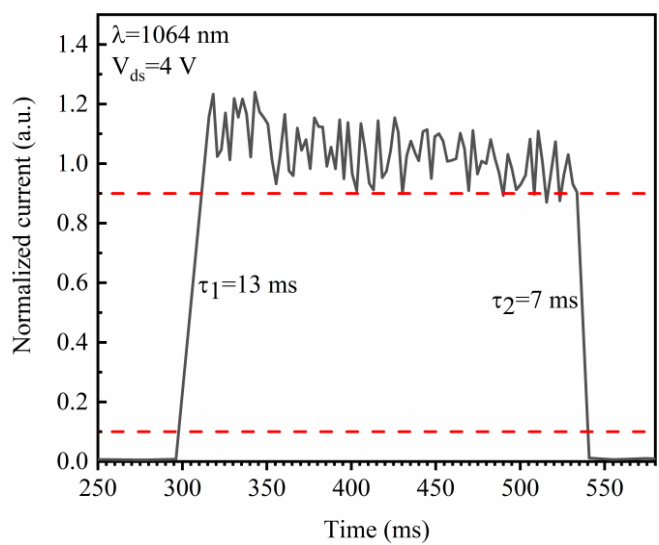

**Figure S8.** Response time curve of the heterostructure under 1064 nm laser illumination.

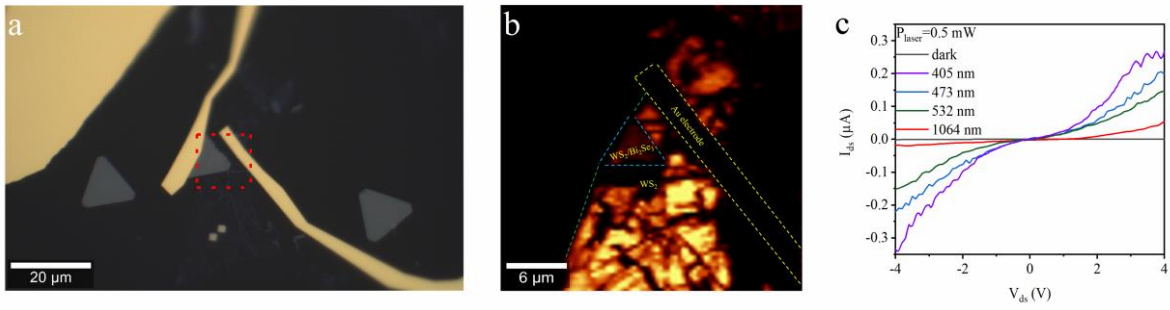

**Figure S9.** Optical image of the WS$_2$/Bi$_2$Se$_3$ heterostructure device and its photodetection performance. (a) Optical image of the heterostructure. (b) PL mapping of the region surrounded by red dashed lines in (a). (c) Broadband response of the device under different wavelength laser illumination at fixed laser power.

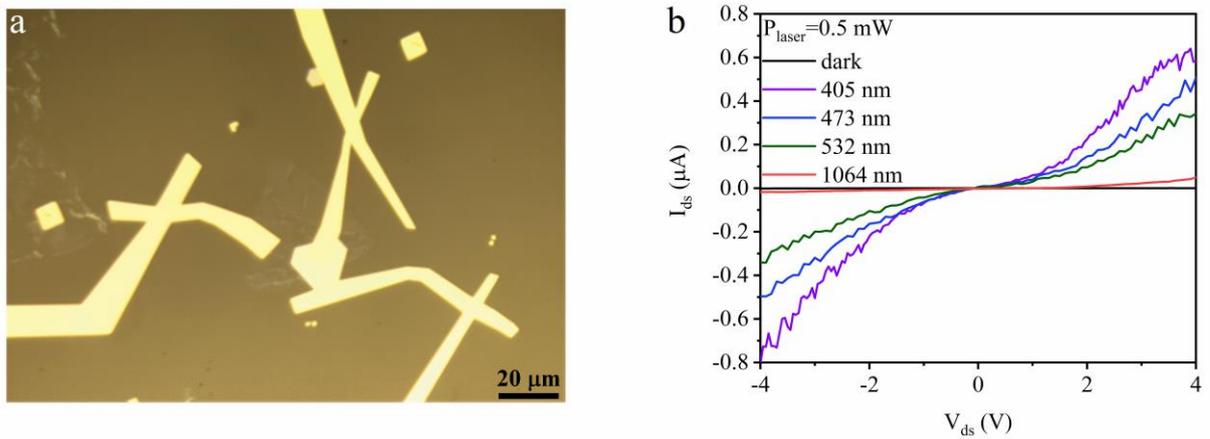

**Figure S10.** Optical image of the WS$_2$/Bi$_2$Se$_3$ heterostructure device and its photodetection performance. (a) Optical image of the heterostructure. (b) Broadband response of the device under different wavelength laser illumination at fixed laser power.